\documentclass[10pt,a4paper]{article}

\pdfoutput=1

\usepackage[english,german]{babel}

\usepackage[longnamesfirst]{natbib} % option round for round paranthesis
\usepackage{multirow}
\usepackage{array}
\usepackage{exscale}
\usepackage{color}
\usepackage{graphicx}

\usepackage{float}

\usepackage[T1]{fontenc}         % for special characters e.g.\ �,�,�
\usepackage[ansinew]{inputenc}   % for special characters e.g.\ �,�,�
\usepackage{ae}

\usepackage{url}
\newcommand{\urlBiBTeX}[1]{\url{#1}}

\usepackage[centertags]{amsmath}
\usepackage{amsfonts}
\usepackage{amssymb}
\usepackage{amsthm}
\usepackage{mathrsfs}
\numberwithin{equation}{section} % for equation numbers like (1.1)

\theoremstyle{plain}

\theoremstyle{definition}

\theoremstyle{remark}

\providecommand{\secref}[1]{Section~\ref{#1}}

\providecommand{\subsubsecref}[1]{Section~\ref{#1}}

\providecommand{\figref}[1]{Figure~\ref{#1}}

\def\D{\dif\!}
\def\tT{\ensuremath{{t\in\left[0,T\right]}}}

\DeclareMathOperator{\dif}{d}
\DeclareMathOperator{\Ex}{E}

\title{On Calibrating Stochastic Volatility Models with time-dependent Parameters\footnote{With best thanks to C\'{e}dric de Masson d'Autume who contributed some of the code used in the numerical example.}}
\author{Wolfgang Putsch\"ogl\footnote{\texttt{wolfgang.putschoegl@unicreditgroup.at}, Counterparty Risk Management \& Analysis, UniCredit Bank Austria, A-1090 Vienna.
\newline 
\textit{Disclaimer: The opinion expressed here is that of the author and does not reflect the views or policies of his
employer.} }}
\begin{document}

\typeout{#####################################}

\selectlanguage{english}

% TITLE ------------------------------------------------------------------
\maketitle

\begin{abstract}
We consider stochastic volatility models using piecewise constant parameters. We suggest a hybrid optimization algorithm for fitting the models to a volatility surface and provide some numerical results. Finally, we provide an outlook on how to further improve the calibration procedure.
\end{abstract}

% INPUT ------------------------------------------------------------------

\section{Introduction}

Over the past two decades various stochastic volatility models have been introduced to explain stylized facts observed in the market. 
Some of the most popular models are the Heston stochastic volatility model and an extension allowing for jumps, cf.\ 
\cite{He93} and \cite{Ba96}, respectively, the Variance-Gamma model introduced by \cite{MCC98}, the model introduced by \cite{BNS01} and L\'{e}vy models with stochastic time, see \cite{CGHD03}.
 
There are basically two approaches to calibrate a stochastic volatility model. It is possible to calibrate the model to historical data or to (liquid) present option prices. For the first approach there exist various methods like maximum likelihood methods, efficient method of moments, or Markov Chain Monte Carlo methods, see \cite{AK06,GT96} and \cite{FS07}, respectively, and the references therein. For various applications the second approach is favorable since the fitted model explains current prices observed on the market, hence we will subsequently consider the latter approach. 

For fitting the model we use a hybrid optimization algorithm consisting of a genetic algorithm to roughly locate the minimum and subsequently a pattern search algorithm with the objective to refine the result. This algorithm allows to take nonlinear constraints like the Feller condition for the Heston model into account. 
Evolutionary algorithms like the genetic algorithm have already been used for calibration e.g.\ by \cite{HC05}. 
%After fitting the model we may use the fitted model to valuate products that are not traded liquidly on the market.

The outline of this paper is as follows. We start in \secref{sec:Models} with an introduction to different models for option pricing. We outline in \secref{sec:FittingtheModel} how to fit a stochastic volatility model to a given volatility surface and give our conclusions in \secref{sec:Conclusion}.

% \bigskip
% \noindent
% \emph{Notation.}
% The symbol $^\top$ will denote transposition. For a vector $v$, $\Diag(v)$ is the diagonal matrix with diagonal $v$.
% For a matrix $M$, $\diag(M)$ is the vector consisting of the diagonal of the matrix $M$.
% We use the symbol $\Unit_n$ for the $n$-dimensional vector whose entries all equal 1. The symbol $\Unit_{n\times d}$ denotes the $(n\times d)$-dimensional matrix whose entries all equal $1$.
% We write $\char_{A}(\omega) $ for the characteristic function which is 1 for $\omega\in A$ and 0 else.
% The symbol $\Id_n$ denotes  the $n$-dimensional identity matrix. Moreover, $\Fc^X=(\Fc_t^X)_{t\in\left[0,T\right]}$ stands for the filtration of augmented $\sigma$-algebras generated by the $\Fc$-adapted process $X=(X_t)_{t\in\left[0,T\right]}$.
% We write $x^-$ for the negative part of $x$: $x^-=\max \left\lbrace -x,0\right\rbrace $, and $x^+$ for the positive part of $x$: $x^+=\max \left\lbrace x,0\right\rbrace$.
% %The symbol $\delta_{ij}$ denotes the Kronecker-symbol. We write $\left[\;\cdot\;%\right]_t$ for the quadratic variation up to time $t$.
% We denote the $k$-th component of a vector $a$ by $a^k$. The $k$-th row and column of a matrix $A$ are denoted by $(A)^{k\;\fatdot}$ and $(A)^{\fatdot\;k}$, respectively.

\section{Model Setup}
\label{sec:Models}

% We describe two different models to price options: the Heston model without jump and the Heston model with jumps. Let us first define some notations for both models.\\
Let $X=(X_t)_{\tT}$ denote the stock price process and $\phi_t(u)$ the characteristic function of the log-price $\ln(X_t)$, defined by
\begin{equation*}
\phi_t(u)=\Ex\Bigl[\exp (\imath u\ln (X_t))|X_0,v_0\Bigr]\;.
\end{equation*}
For ease of notation we assume the risk-free interest rate $r_t\equiv r$ to be constant.

\subsection{The Heston Model}
\label{subsec:TheHestonmodel}

\cite{He93} introduced a stochastic volatility model which satisfies a lot of stylized facts like fat tails, volatility clustering, skewness etc., cf.~\cite{Co01}. The Heston model is defined under the risk neutral measure by the coupled two-dimensional SDE
\begin{align}
\label{eq:SDE_Heston}
\D X_t&=rX_t\D t+ \sqrt{v_t}X_t\D W_t\;,\\
\label{eq:SDE_Heston2}
\D v_t&=\kappa(\theta-v_t)\D t+ \sigma\sqrt{v_t}\D \widetilde{W}_t\;,
\end{align} 
where $W=(W_t)_{\tT}$ and $\widetilde{W}=(\widetilde{W}_t)_{\tT}$ are one-dimensional Brownian motions with correlation $\rho$ and $v=(v_t)_{\tT}$ denotes the variance process. The involved parameters are the rate of mean reversion $\kappa > 0$, the long term variance $\theta > 0$, the volatility of variance $\sigma > 0$ (often called the vol of vol), the correlation $-1 < \rho <1$, and the initial variance $v_0$. If the Feller condition $2\kappa\theta > \sigma^{2}$ is satisfied, then the variance process is always positive and cannot reach zero. In absence of the stochastic factor, we have an exponential attraction to long term variance, the equilibrium point being $v_t = \theta$. Typically, the correlation $\rho$ is negative, hence a down-move in the stock price is positively correlated with an up-move in the volatility.

\subsubsection{Time-Independent Parameters}
\label{subsubsec:TimeIndependentParameters}

Note that there are two representations of the characteristic function. The first one used e.g.\ by \cite{He93} or \cite{KJ05} reads
\begin{align*}
\phi_t^{(1)}(u)&= \exp (\imath u(\ln X_0 + rt))\;  \\ 
&\times \exp (\theta\kappa\sigma^{-2}((\kappa-\rho\sigma \imath u+d)t-2 \ln((1-g_1e^{dt})/(1-g_1))))\;  \\ 
&\times \exp (v_0\sigma^{-2}(\kappa-\rho\sigma\imath u+d)(1-e^{dt})/(1-g_1e^{dt}))\;,  
\end{align*}
and the second one e.g.\ used by \cite {SST04} and \cite{Ga06} is given by
\begin{align}
\phi_t^{(2)}(u)&= \exp (\imath u(\ln X_0 + rt))\; \label{eq:cf} \\
&\times \exp (\theta\kappa\sigma^{-2}((\kappa-\rho\sigma \imath u-d)t-2 \ln((1-g_2e^{dt})/(1-g_2))))\; \notag \\
&\times \exp (v_0\sigma^{-2}(\kappa-\rho\sigma\imath u-d)(1-e^{dt})/(1-g_2e^{dt}))\;, \notag
\end{align}
where
\begin{align}
d&=\sqrt{(\rho\sigma u\imath-\kappa)^{2}+\sigma^{2}(\imath u+u^2)}\;,\label{eq:d}\\
g_1&=\frac{\kappa-\rho\sigma\imath u+d}{\kappa-\rho\sigma\imath u-d}\;,  \qquad
g_2=\frac{\kappa-\rho\sigma\imath u-d}{\kappa-\rho\sigma\imath u+d}=\frac{1}{g_1}\;\label{eq:g_2}.
\end{align}
\cite{AMST07} prove in a detailed analysis that $\phi^{(1)}$ is unstable under certain conditions whereas $\phi^{(2)}$ is stable under the full parameter space.

\subsubsection{Piecewise Constant Parameters}
\label{subsubsec:PiecewiseConstantParameters}

To obtain a good fit for a volatility surface, i.e., for several maturities at once, is fairly difficult to obtain using a stationary process. One way to obtain a good fit is to use processes for the price process having piecewise constant parameters. The solution for the characteristic function for piecewise constant parameters is given by \cite{MN03} who derived it using a Computer Algebra system. The full derivation of the solution is provided by \cite{E07} who also presents a formula which is in line with the findings of \cite{AMST07}. The solution is given by
\begin{align*}
\phi_t(u)&=\exp \bigl(C+Dv_0+\imath u\ln (X_0)\bigr)\;,
\end{align*}
where
\begin{align*}
C&=\imath t  ru +\frac{\kappa\theta}{\sigma^2}\Bigl(-2\ln \Bigl(\frac{1-\tilde{g}e^{-dt}}{1-\tilde{g}}\Bigr)+(\kappa-\rho\sigma\imath u-d) t \Bigr)+C^0\;,\\
D&=\frac{\kappa-\rho\sigma u\imath}{\sigma^2}\Bigl(\frac{g-\tilde{g}e^{-d t }}{1-\tilde{g}e^{-d t }}\Bigr)\;,\qquad
\tilde{g}=\frac{\kappa-\rho\sigma\imath u-d-D^0\sigma^2}{\kappa-\rho\sigma\imath u+d-D^0\sigma^2}\;,
\end{align*}
with $d$ as in \eqref{eq:d} and $g = g_2$ as in \eqref{eq:g_2}.

\medskip
The solution is close to the Heston one \eqref{eq:cf}. The time interval to maturity $[0,T]$ is divided into $n$ sub-intervals $[0,t_1],\dots,[t_k,t_{k+1}],\dots,[t_{n-1},T]$ where $t_k,k=1,\dots,n-1$ is the time of model parameter jumps. Model parameters are constant during $[t_k,t_{k+1}]$ but different for each sub interval. The initial condition for the first sub interval from the end $[0,t_1]$ is zero. For the second sub interval $[t_1,t_2]$ we use $C$ and $D$ from the first sub interval for $C^0$ and $D^0$, the initial conditions. The same procedure is repeated at each time $t_k,k=2,\dots,n-1$ of the parameters jumps.

\subsection{The Heston Model with Jumps}

\cite{Ba96} extended the Heston model by introducing jumps in the asset price. The coupled two-dimensional SDE \eqref{eq:SDE_Heston}--\eqref{eq:SDE_Heston2} becomes
\begin{align*}
\label{eq:SDE_Bates}
\D X_t&=(r-\lambda\mu_J)X_t\D t+ \sqrt{v_t}X_t\D W_t+J_t\D N_t\;,\\
\D v_t&=\kappa(\theta-v_t)\D t+ \sigma\sqrt{v_t}\D \widetilde{W}_t\;, 
\end{align*}
with the same notations as in \eqref{eq:SDE_Heston} and \eqref{eq:SDE_Heston2} and with $N=(N_t)_{\tT}$ being an independent Poisson process with intensity $\lambda>0$ (i.e., $\Ex[N_t]=\lambda t$) and $J=(J_t)_{\tT}$ is the percentage jump size that is log-normally i.i.d.\ over time with mean $\mu_J$. The standard deviation of $\ln(1+J_t)$ is $\sigma_J$ and
\begin{align*}
\ln(1+J_t)\sim \mathcal{N}\bigl(\ln (1+\mu_J)-\frac{\sigma_J^2}{2},\sigma_J^2\bigr)\;. 
\end{align*}

\subsubsection{Time-Independent Parameters}
\label{subsubsec:TimeIndependentParametersBates}

Since the jumps and the diffusion part are independent, the characteristic function for $\ln(X_t)$ is defined by
\begin{align*}
\phi_t(u)=\phi_t^D(u)\phi_t^J(u)\;,
\end{align*}
where $\phi_t^D\equiv \phi_t^{(2)}$ as in \eqref{eq:cf} is the characteristic function for the diffusion part and $\phi_t^J(u)$ the characteristic function corresponding to the jump part is given by
\begin{align*}
\phi_t^J(u)=\exp\Bigl(-\lambda\mu_J\imath ut+\lambda t\bigl((1+\mu_J)^{\imath u}\exp(\sigma_J^2(\imath u/2)(\imath u-1))-1\bigr)\Bigr)\;.
\end{align*}

\subsubsection{Piecewise Constant Parameters}
\label{subsubsec:PiecewiseConstantParametersBates}

We still define the characteristic function as $\phi_t(u)=\phi_t^D(u)\phi_t^J(u)$ with $\phi_t^D$ corresponding to  $\phi_t$ as in \subsubsecref{subsubsec:PiecewiseConstantParameters} and $\phi_t^J$ given by
\begin{align*}
\phi_t^J(u)=&\Ex\Bigl[\exp\Bigl(\imath u\int_0^t{J_s}\D N_s\Bigr)\Bigr]\;
%=&\Ex\Bigl[\exp\Bigl(\imath u\int_0^{t_1}{J_s}\D N_s+\imath u\int_{t_1}^{t_2}{J_s}\D N_s+\dots+\imath u\int_{t_{N-1}}^{t_N}{J_s}\D N_s\Bigr)\Bigr]\;\\
=\Ex\Bigl[\exp\Bigl(\imath u\sum_{k=0}^{N-1}\int_{t_k}^{t_{k+1}}{J_s}\D N_s\Bigr)\Bigr]\;\\
=&\prod_{k=0}^{N-1}{\Ex\Bigl[\exp\Bigl(\imath u\int_{t_k}^{t_{k+1}}{J_s}\D N_s\Bigr)\Bigr]} \;\\
=&\prod_{k=0}^{N-1}\exp\Bigl(-\lambda^{(k)}\mu_J^{(k)}\imath u\Delta_{t_k}+\lambda^{(k)} \Delta_{t_k}((1+\mu_J^{(k)})^{\imath u} \\
& \qquad \times \exp(\sigma_J^{(k)^2}(\imath u/2)(\imath u-1))-1)\Bigr)\;,
\end{align*}
where $\lambda^{(k)}$ and $\mu_J^{(k)}$ are the parameters of the jumps during the sub interval $[t_k,t_{k+1}]$ and $\Delta_{t_k}=t_{k+1}-t_k$.

\subsection{The Variance Gamma Process}
\label{subsec:VGProcess}

The Variance-Gamma (VG) process is obtained by evaluating Brownian motion (with constant drift $\theta$ and volatility $\sigma$) at a random time change given by a Gamma process. Let
\begin{align*}
b_t(\theta,\sigma)=\theta t+\sigma W_t\;,
\end{align*}
where $W=(W_t)_{\tT}$ is a Brownian motion. Then $b_t( \theta, \sigma)$ is a Brownian motion with drift $\theta$ and volatility $\sigma$.

The Gamma process $\gamma_t(\mu,\nu)$ with mean rate $\mu$ and variance rate $\nu$ is a process of independent Gamma increments over time intervals $[t,t+h]$. The VG process $G_t(\sigma, \nu, \theta)$ is then defined by
\begin{align*}
G_t(\sigma,\nu,\theta)=b_{\gamma_t( 1, \nu)}(\theta, \sigma)\;.
\end{align*}
The involved parameters are the volatility of the Brownian motion $\sigma$, the variance rate time change $\nu$ and the drift in the Brownian motion $\theta$. We can control the skewness of the distribution via $\theta$ and the kurtosis with $\nu$. The resulting risk-neutral process for the stock price is
\begin{align*}
X_t=X_0\exp(rt+G_t(\sigma,\theta,\nu)+\omega t)\;,
\end{align*}
where $\omega=(1/\nu)\ln(1-\theta\nu-\sigma^2\nu/2)$.

\subsubsection{Time-Independent Parameters}
\label{subsubsec:TimeIndependentParametersVG}

The characteristic function of $\ln(X_t)$ reads
\begin{align*}
\phi_t(u)=\exp\bigl(\ln(X_0+(r+\omega)t)\bigr)\bigl(1-\imath \theta\nu u+\sigma^2 u^2\nu/2\bigr)^{-t/\nu}\;.
\end{align*}
From this formula, one can obtain a closed formula for both the density function and the option price, for details we refer to \cite{MCC98}.

\subsubsection{Piecewise Constant Parameters}
\label{subsubsec:PiecewiseConstantParametersVG}

For piecewise constant parameters we can use the same method as in \subsubsecref{subsubsec:PiecewiseConstantParametersBates}. For readability we use the notation $G_t$ instead of $G_t(\sigma,\theta,\nu)$, $\Delta G_{t_k}=G_{t_{k+1}}-G_{t_k}$ and $t_N=t$. Note that $G_0=0$ and that the increments $\Delta G_{t_k}$ are independent. Then $\phi_t(u)$ is given by
\begin{align*}
&\phi_t(u)=\Ex\Bigl[\exp(\imath u\ln(X_t))\Bigr]\;\\
%&=\Ex\Bigl[\exp\Bigl(\imath u\ln(X_0\exp(rt+G_t+\omega t))\Bigr)\Bigr]\;\\
&=\Ex\Bigl[\exp\Bigl(\imath u(\ln(X_0)+(r+\omega)t)\Bigr)\exp\bigl(\imath uG_t\bigr)\Bigr]\;\\
%&=\exp\Bigl(\imath u(\ln(X_0)+(r+\omega)t)\Bigr)\Ex\Bigl[\exp\Bigl(\imath uG_t\Bigr)\Bigr]\;\\
&=\exp\Bigl(\imath u(\ln(X_0)+(r+\omega)t)\Bigr)\Ex\Bigl[\prod_{k=0}^{N-1}{\exp\Bigl(\imath u(\Delta G_{t_k})\Bigr)}\Bigr]\; \\
&=\exp\Bigl(\imath u(\ln(X_0)+(r+\omega)t)\Bigr)\prod_{k=0}^{N-1}{\Ex\Bigl[\exp\Bigl(\imath u(\Delta G_{t_k})\Bigr)\Bigr]}\; \\
&=\exp\Bigl(\imath u(\ln(X_0)+(r+\omega)t)\Bigr)\prod_{k=0}^{N-1}{\Bigl(1-\imath \theta^{(k)}\nu^{(k)} u+\sigma^{(k)^2} u^2\nu^{(k)}/2\Bigr)^{-\Delta_{t_k}/\nu^{(k)}}}\;,
\end{align*}
where $\theta^{(k)}$, $\sigma^{(k)}$ and $\nu^{(k)}$ are the parameters of the models during the sub interval $[t_k,t_{k+1}]$ and $\Delta_{t_k}=t_{k+1}-t_k$.

\section{Fitting the Model}
\label{sec:FittingtheModel}

To fit a stochastic volatility model to a given volatility surface, we first have to compute option prices, then to compute the implied volatility from these prices and finally to resolve the Optimization Problem.

% regularization, weights -> vega weighted, or on reliability, atm bigger?

\subsection{Optimization Problem}
\label{subsec:Optimization_Problem}

We assume a set of option prices for different maturities and strikes as given.
To calibrate a stochastic volatility model, we have to minimize an objective function. We strive to minimize the mean squared error (MSE) of  volatilities, i.e.,
\begin{align*}
\text{minimize}\sum_{\text{Options}}(\sigma_{\text{given}}-\sigma_{\text{fit}})^2\;,
\end{align*}
where $\sigma_{\text{given}}$ is a given vector of market Black-Scholes implied volatilities  and $\sigma_{\text{fit}}$ are the implied Black-Scholes volatility resulting from pricing the option with a stochastic volatility model. 
Instead of fitting the implied volatilities we could also fit option prices. Further it is possible to weight the differences between the given and the fitted prices. The choice of the weights may depend on the liquidity of a given option which can be assessed from the bid-ask spread. \cite[Chapter~13]{Co04} suggest to weight the option prices by squared Black-Scholes vegas evaluated at the implied volatilities of the market option prices. 
If we have a prior distribution, e.g.\ based on the posterior distribution from parameter estimation from historical data  based on a Markov Chain Monte Carlo method, we may apply regularization techniques for which we refer to \cite[Chapter~13]{Co04}.

\subsection{Computation of Option prices}
\label{subsec:Computation_of_Option_prices}

For our application it is crucial to have an efficient method to price plain vanilla options. Therefore we use the Fast Fourier Transform (FFT) introduced for option pricing by \cite{CM99}. The basic idea of this method is to develop an analytical expression for the Fourier transform of the option price and then to get the price back by Fourier inversion. Assuming no dividends and constant interest rates $r$, the initial value $C_0$ of a plain vanilla European call option is determined as
\begin{align*}
C_0=X_0\Pi_1-Ke^{-rT}\Pi_2\;,
\end{align*}
where
\begin{align*}
\Pi_1=\frac{1}{2}+\frac{1}{\pi}\int_0^{+\infty} {\mathop{\mathrm{Re}}\Bigl[\frac{e^{-\imath u\ln(K)}\phi_T(u-\imath)}{\imath u\phi_T(-\imath)}\Bigr]}\D u\;,\\
\Pi_2=\frac{1}{2}+\frac{1}{\pi}\int_0^{+\infty} {\mathop{\mathrm{Re}}\Bigl[\frac{e^{-\imath u\ln(K)}\phi_T(u)}{\imath u}\Bigr]}\D u\;,
\end{align*}
with $\phi_T(u)$ being the characteristic function of $\ln(X_T)$. Corresponding plain vanilla European put option prices can be derived via the Put-Call-parity equation.
Since we cannot evaluate the integral using the FFT, we rather use the expression
\begin{align*}
C_T(k)=\int_k^{+\infty}e^{-rT}(e^s-e^k)q_T(s) \D s\;,
\end{align*}
where $k$ denotes the log of the strike $K$ and $q_T$ the risk-neutral density of the log price of the underlying. Since the call pricing function is not square integrable, we consider the modified call price $c_T(k)$ defined by
\begin{align*}
c_T(k)=\exp(\alpha k)C_T(k)\;,
\end{align*}
for a suitable $\alpha>0$. We consider now the Fourier transform of $c_T(k)$ given by
\begin{align*}
\psi_T(v)=\int_{-\infty}^{+\infty}e^{\imath vk}c_T(k)\D k=\frac{\exp(-rT)\phi_T(v-(\alpha+1)\imath)}{\alpha^2+\alpha-v^2+\imath(2\alpha+1)v}\;.
\end{align*}
Then we can use the inverse transform to obtain $C_T(k)$
\begin{align*}
C_T(k)=\frac{\exp(-\alpha k)}{\pi}\int_0^{+\infty}e^{-\imath vk}\psi_T(v)\D v\;.
\end{align*}
and the FFT to compute an approximation of $C_T(k)$ via
\begin{align*}
C_T(k)\approx \frac{\exp(-\alpha k)}{\pi}\sum_{j=1}^{N}e^{-\imath v_jk}\psi_T(v_j)\eta\;,
\end{align*}
where $\eta$ is the distance of the points of the integration grid and $v_j=\eta(j-1)$. 
For more details we refer to \cite{CM99}. This method is very fast since efficient implementations of the FFT exist. Moreover, this approach allows to compute the option prices for several strikes at once.
Note that this approach only depends on the specific models via the characteristic function $\phi$.
%Then to obtain the option prices, we just have to evaluate $\phi_T(u)$. We can use $\phi_T(u)$ for time-independent parameters or for piecewise constant parameters. 
%The first case is described in \subsubsecref{subsubsec:TimeIndependentParameters} and \ref{subsubsec:TimeIndependentParametersBates} and the second one in \subsubsecref{subsubsec:PiecewiseConstantParameters} and \ref{subsubsec:PiecewiseConstantParametersBates}.

\subsection{Hybrid Optimization Algorithm}

We suggest a \emph{hybrid optimization algorithm} which consists of a genetic algorithm with the purpose of locating a minimum within the specified bounds for the parameters and a subsequent direct search method to refine the solution. For a detailed reference concerning genetic algorithms we refer to \cite{BFM00}.

\subsubsection{Genetic Algorithm (GA)}

\paragraph{Outline of the Algorithm}

The following outline summarizes how the genetic algorithm works:

\begin{enumerate}
\item The algorithm begins by creating randomly an initial population. The population is chosen such that possible constraints (e.g.\ upper or lower bounds on the variables) are satisfied. It is also possible to provide an initial population or to specify only some members of the initial population.
\item The algorithm evaluates the fitness of each individual in the population. 
\item The algorithm then creates iteratively a sequence of populations. At each step, the algorithm creates the next population based on the current generation. Therefore the following steps are performed: 
	\begin{enumerate}
	\item The algorithm evaluates the \emph{fitness} of each individual in the population. The fitness is a measure on how well the optimization problem is solved.
	\item A certain number of individuals in the current population (called the \emph{Elite count}) that have the highest fitness values are chosen as elite children and are passed on to the next population. 
	\item A certain number of children of the next generation are produced by combining a pair of parents (crossover). The number is specified as fraction of the current population (\emph{Crossover fraction}).
	\item The remaining children are produced by making random changes to a parent (\emph{Mutation}).  
	\item The children of the current population form the next generation.
	\end{enumerate}
\item The algorithm runs until one of the stopping criteria is met (e.g.\ maximum number of (stall) generations/time reached, average change in fitness function less than a certain value, certain fitness value reached) 
\end{enumerate}

For the Heston and Bates model we sample an initial population uniformly distributed between lower and upper bounds on the variables such that they satisfy the Feller condition. If parameter estimates from historical data are available these parameters may be included in the initial population. If the parameter estimates are based on a Markov Chain Monte Carlo method such that we get a posterior distribution for each parameter we may sample the initial population from the posterior distributions and the bounds on the parameters may be derived from the posterior distribution as well, e.g.\ as a quantile of the distribution.

\smallskip

We use a \emph{crossover} method that creates children as a linear combination of two parents, i.e., that lie on the line containing the two parents. We may choose the child randomly on the line between the two parents or alternatively
 a small distance away from the parent with the better fitness value in the direction away from the parent with the worse fitness value. Note that for the Heston model if parent$_1$ and parent$_2$ with parameters $(\theta_i,\kappa_i,\rho_i,\sigma_i,\nu_{0,i})$  satisfy the Feller condition $2\kappa_i\theta_i-\sigma_i^2>0$ for $i=1,2$, then the Feller condition is also satisfied for a parameter set on the line between the two parents since 
\begin{equation*}
2(\lambda\kappa_1+(1-\lambda)\kappa_2)(\lambda\theta_1+(1-\lambda)\theta_2)-(\lambda\sigma_1+(1-\lambda)\sigma_2)^2> \frac{(\sigma_1\theta_2-\sigma_2\theta_1)^2}{2\theta_1\theta_2}\geq0 
\end{equation*} 
for $\lambda\in(0,1)$. 

\smallskip

For mutation we randomly generate directions that are adaptive with respect to the last successful or unsuccessful generation. A step length is chosen along each direction so that upper and lower bounds as well as the Feller constraint are satisfied. 

\medskip

\subsubsection{Direct Search Method}

After applying the GA with the purpose to locate a minimum in the specified bounds we use a pattern search algorithm to refine the result. We use complete polling to avoid local minima and only poll if the Feller constraint is satisfied.

\subsection{Fitting for Piecewise Constant Parameters}

For piecewise constant parameters we apply a bootstrapping method, i.e., after calibrating to one maturity we calibrate for the subsequent maturity, starting with the first maturity. We use as initial population for calibration the fitted parameters of the previous maturities. 

If the calibration has to be done frequently, e.g.\ daily, we can achieve a fast recalibration by supplying additionally the parameters from the previous run as initial population for the new run (for the respective maturity). 
% ------------------------------------------------------------------------

\subsection{Numerical Example}

In this section we provide numerical results for fitting the Heston model. We consider an FX volatility surface for the Yen/Dollar exchange rate from 21/05/2008 with values $\sigma_{\text{given}}$ for a given set of delta values $\Delta_{\text{given}}$. Since we need the volatility surface in terms of strikes, we transform $\Delta_{\text{given}}$ into $K_{\text{given}}$ using
\begin{align*}
K_{\text{given}}=\frac{S}{\exp\bigl(\mathcal N^{-1}(\Delta_{\text{given}})\sigma\sqrt t-(r+\frac{\sigma^2}{2})t\bigr)}\;,
\end{align*}
where $\mathcal N$ denotes the standard normal cumulative distribution function. Moreover, for at the money options, we have to substitute the strike with the corresponding forward price. To obtain $\sigma_{\text{fit}}$, we compute the option prices  and retrieve the corresponding Black-Scholes model implied volatilities $\sigma_{\text{fit}}$ via a standard bisection method.
For the computation of option prices we use the FFT as described in \secref{subsec:Computation_of_Option_prices}.
We use $\alpha=0.75$  since \cite{SST04} show in an empirical study that with this value prices are well replicated for many model parameters. The parameters $N=2^{16}$ and $\eta=0.015$ turned out to be a reasonable trade-off between accuracy and speed.

\begin{figure}
\centering

\begin{minipage}[b]{0.45\textwidth}
\centering
\includegraphics[width=\textwidth]{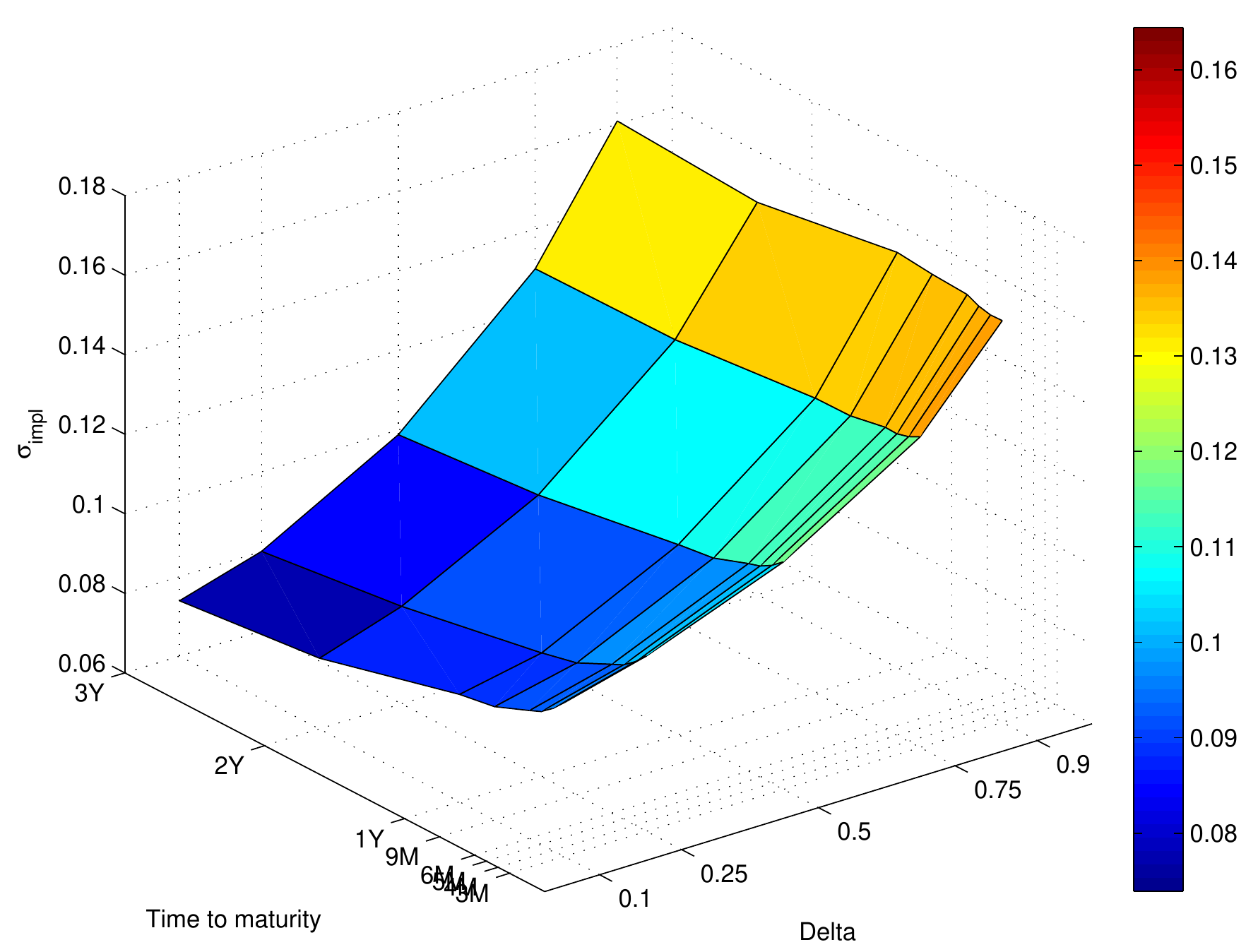}
\caption{given volatility surface}
\label{fig:Heston_impliedVola_given}
\end{minipage}%
\hspace{0.04\textwidth}%
\begin{minipage}[b]{0.45\textwidth}
\centering
\includegraphics[width=\textwidth]{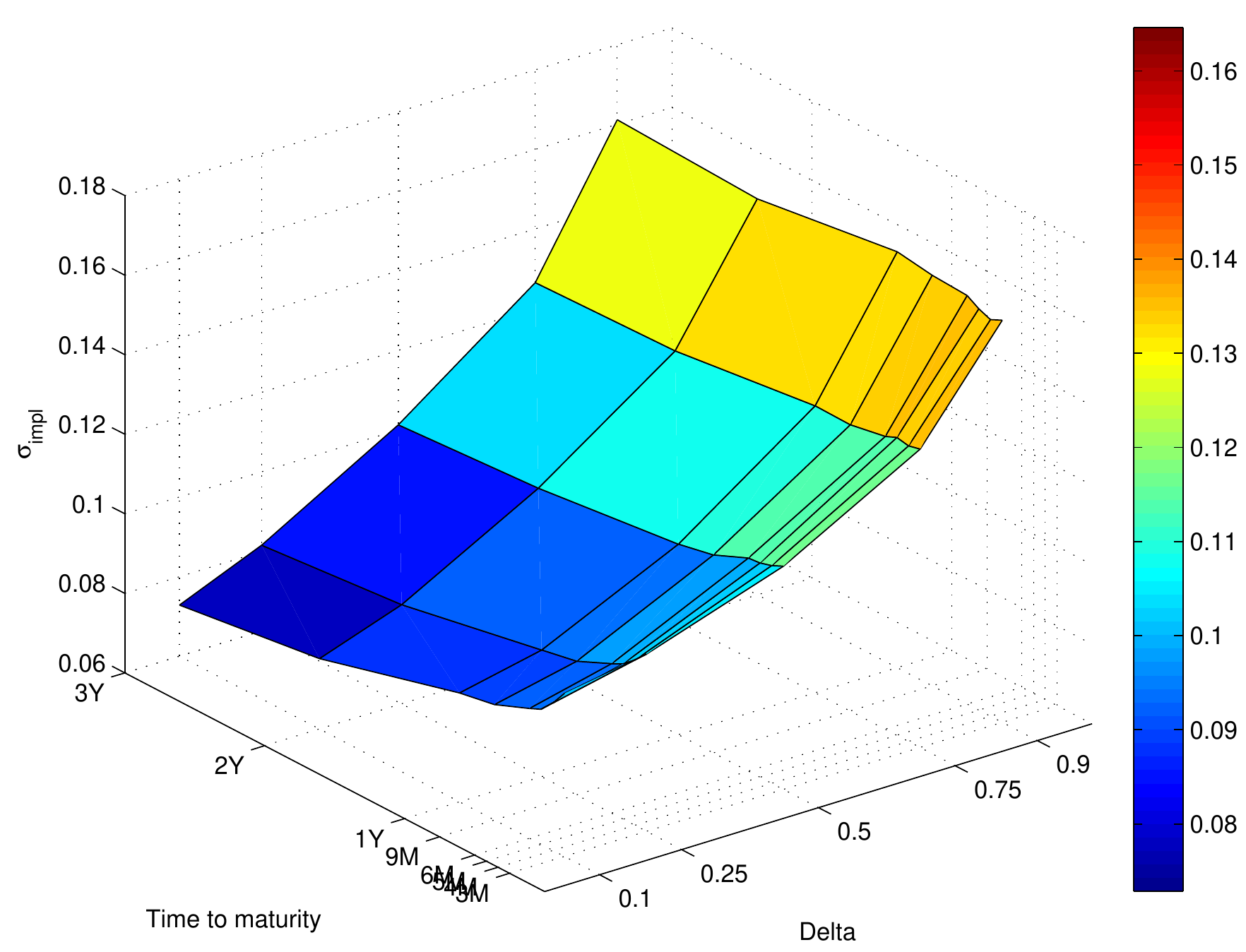}
\caption{fitted volatility surface}
\label{fig:Heston_impliedVola_fitted}
\end{minipage}\\[00pt]
\end{figure}

\begin{figure}
\centering

\begin{minipage}[b]{\textwidth}
\centering
\includegraphics[width=0.6\textwidth]{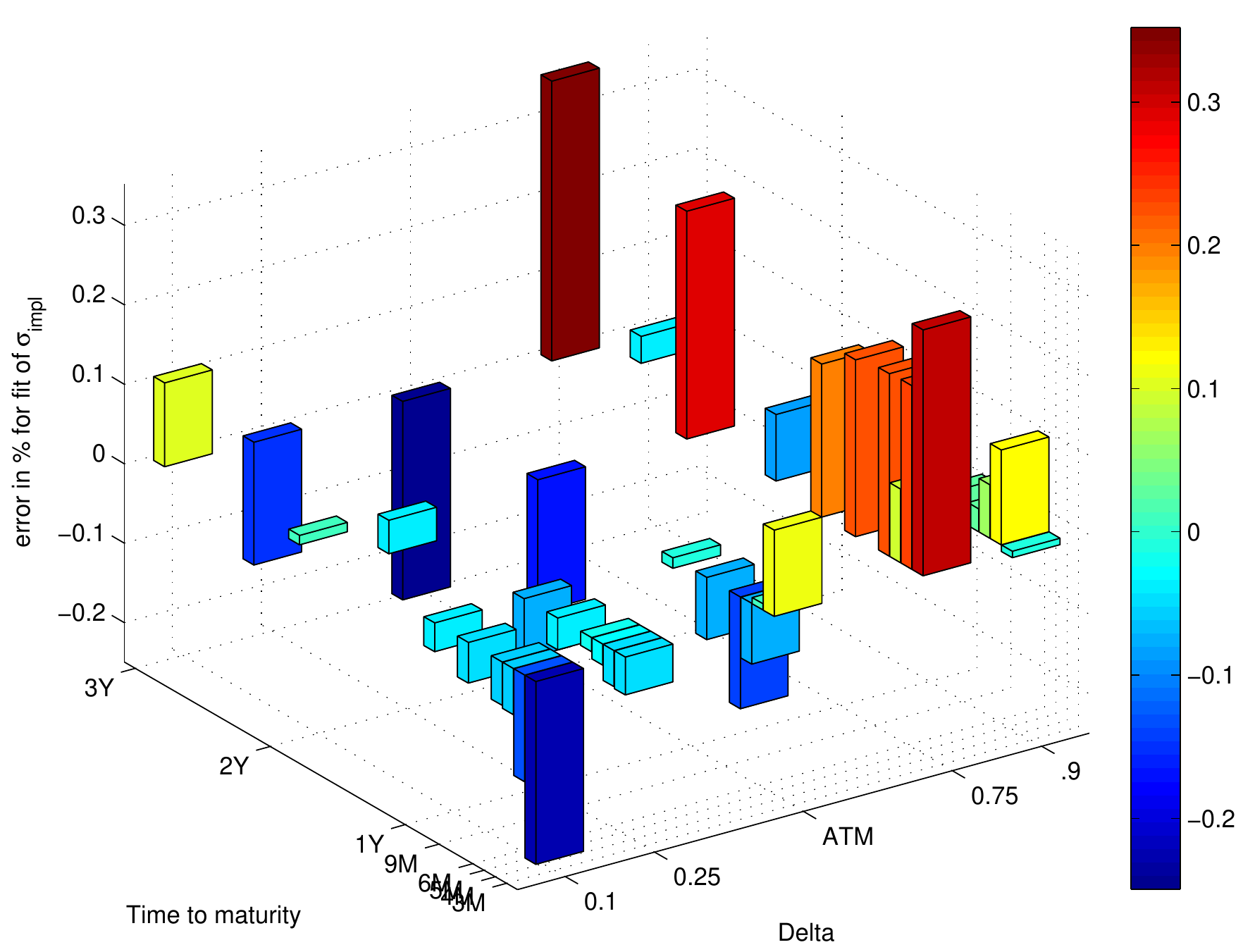}
\\[-12pt]\caption{error (in \%)}
\label{fig:Heston_impliedVola_error}
\end{minipage}
\end{figure}

\begin{figure}
\centering
\begin{minipage}[b]{0.7\textwidth}
\centering
\includegraphics[width=\textwidth]{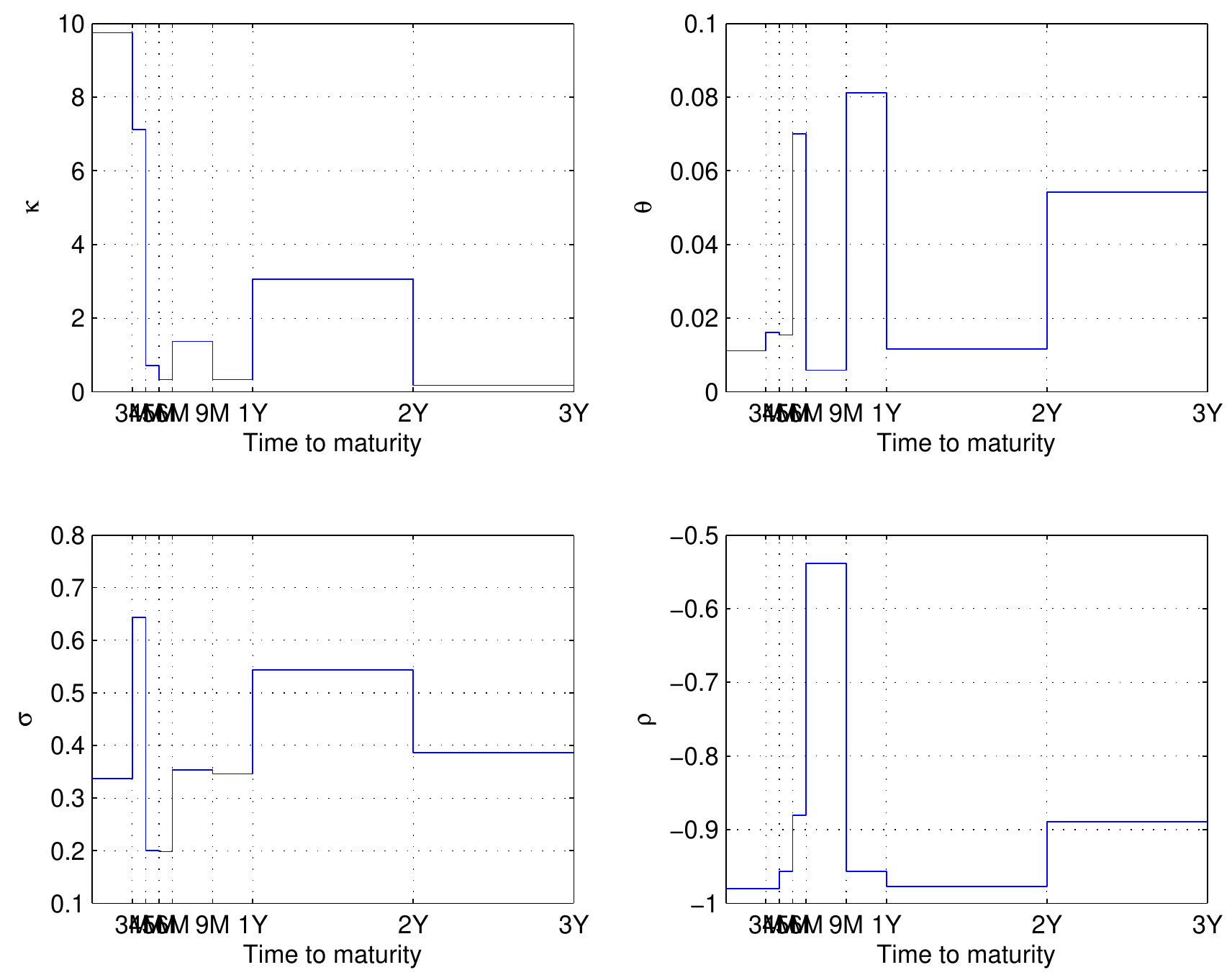}
\\[-12pt]\caption{(time dependent) parameters}
\label{fig:Heston_time_dependent_params}
\end{minipage}

\end{figure}

\begin{figure}
\centering

\begin{minipage}[b]{0.45\textwidth}
\centering
\includegraphics[width=\textwidth]{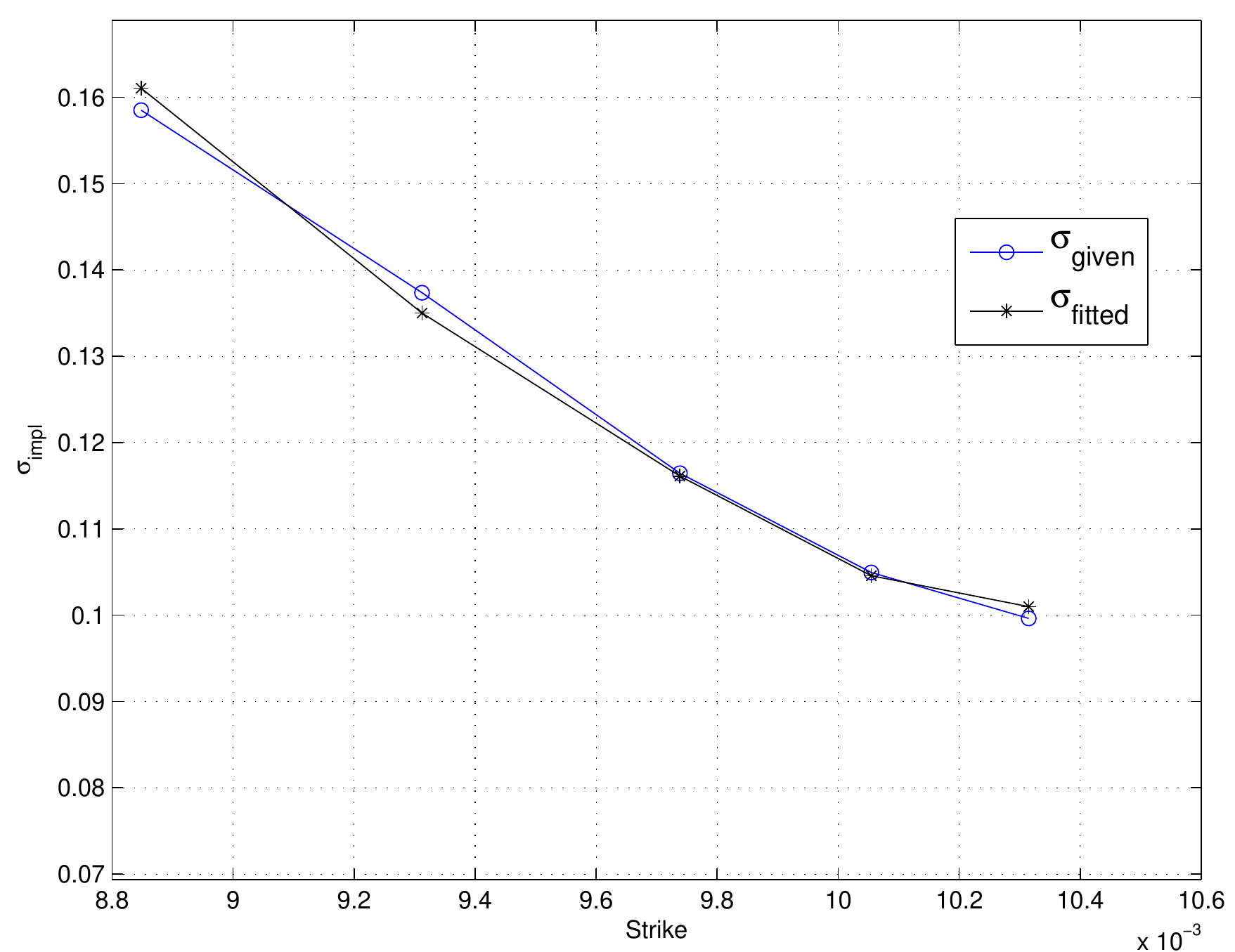}
\\[-12pt]\caption{fit for maturity 3M}
\label{fig:Heston_impliedVola_error_T_3M}
\end{minipage}%
\hspace{0.04\textwidth}%
\begin{minipage}[b]{0.45\textwidth}
\centering
\includegraphics[width=\textwidth]{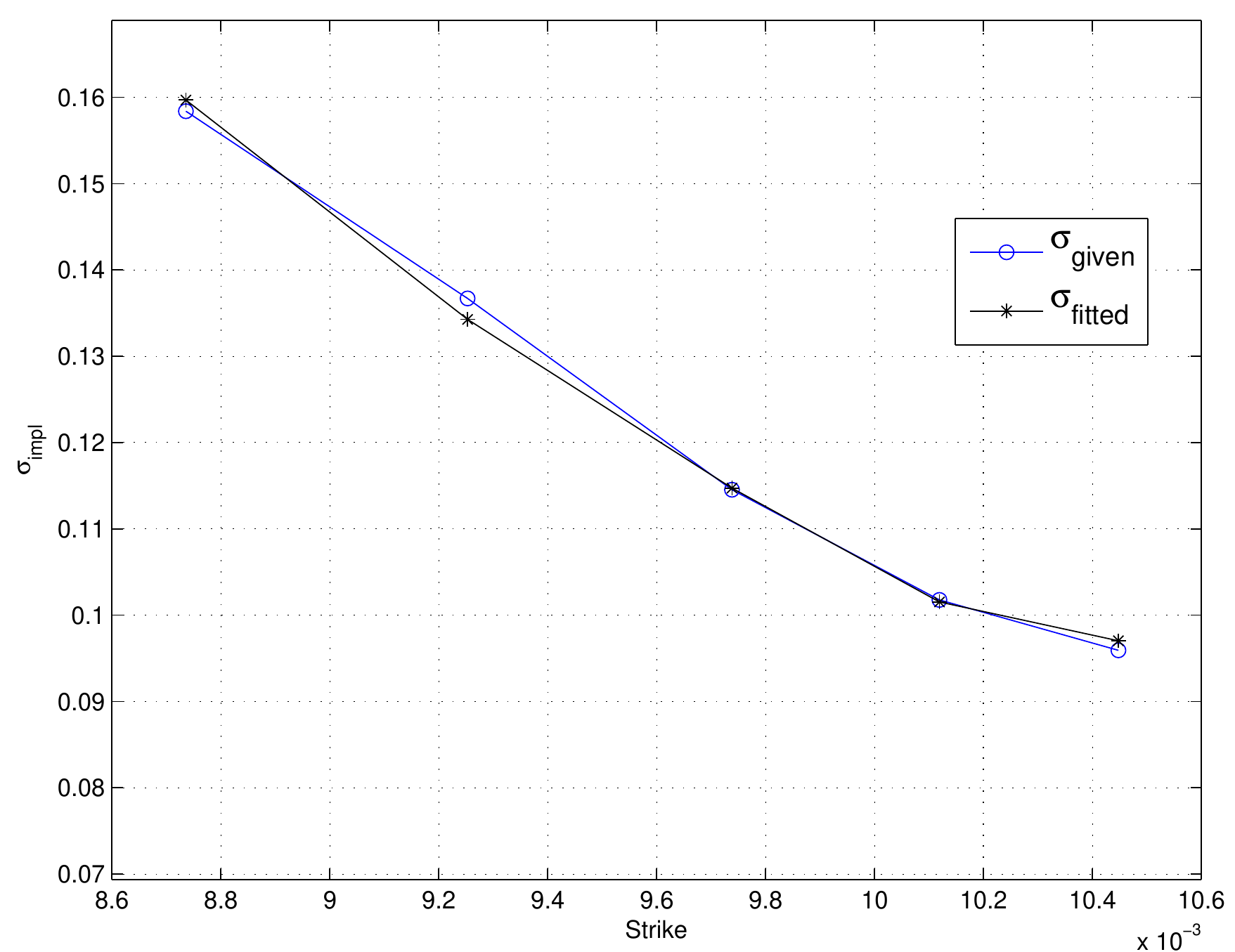}
\\[-12pt]\caption{fit for maturity 4M}
\label{fig:Heston_impliedVola_error_T_4M}
\end{minipage}\\[20pt]

\begin{minipage}[b]{0.45\textwidth}
\centering
\includegraphics[width=\textwidth]{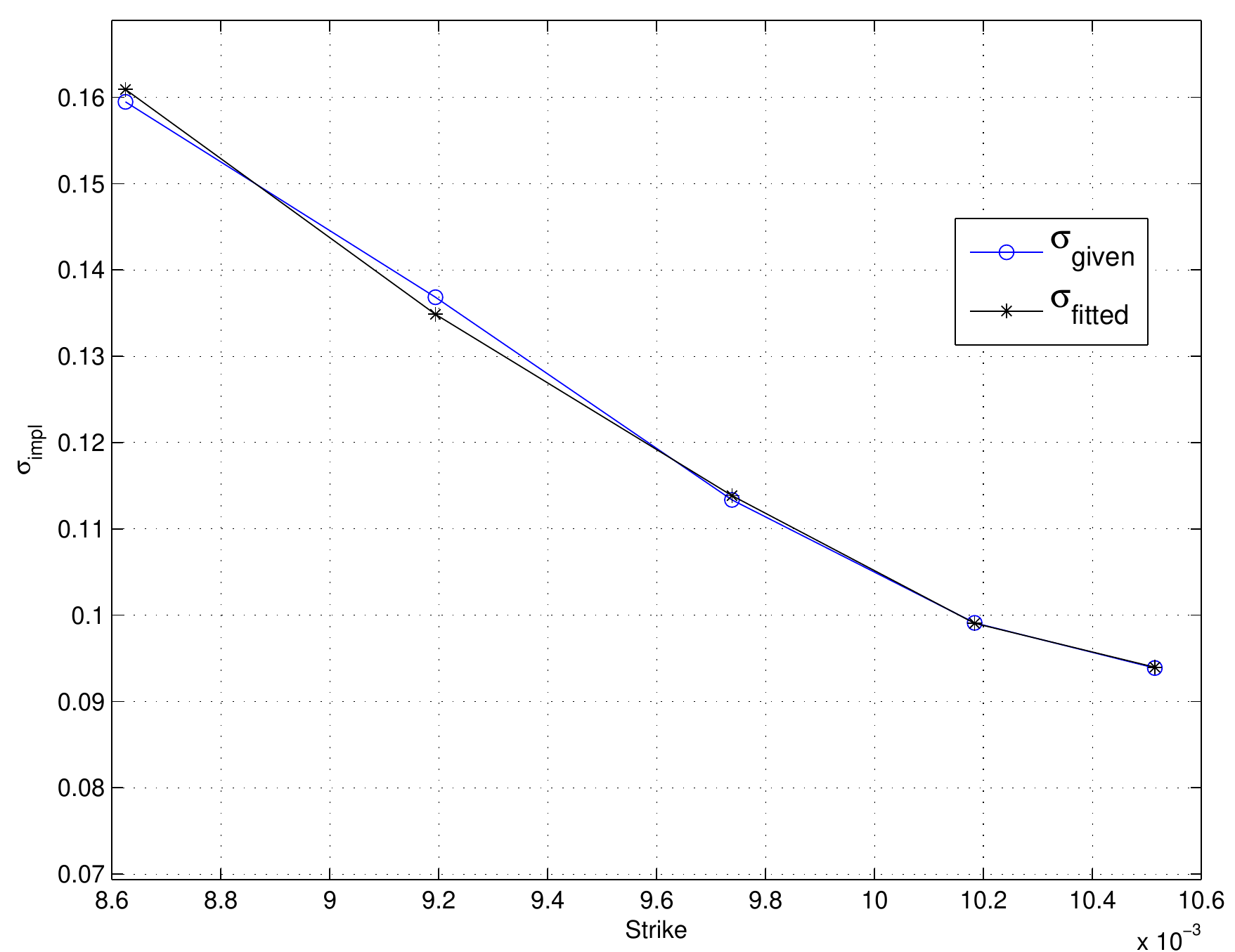}
\\[-12pt]\caption{fit for maturity 5M}
\label{fig:Heston_impliedVola_error_T_5M}
\end{minipage}%
\hspace{0.04\textwidth}%
\begin{minipage}[b]{0.45\textwidth}
\centering
\includegraphics[width=\textwidth]{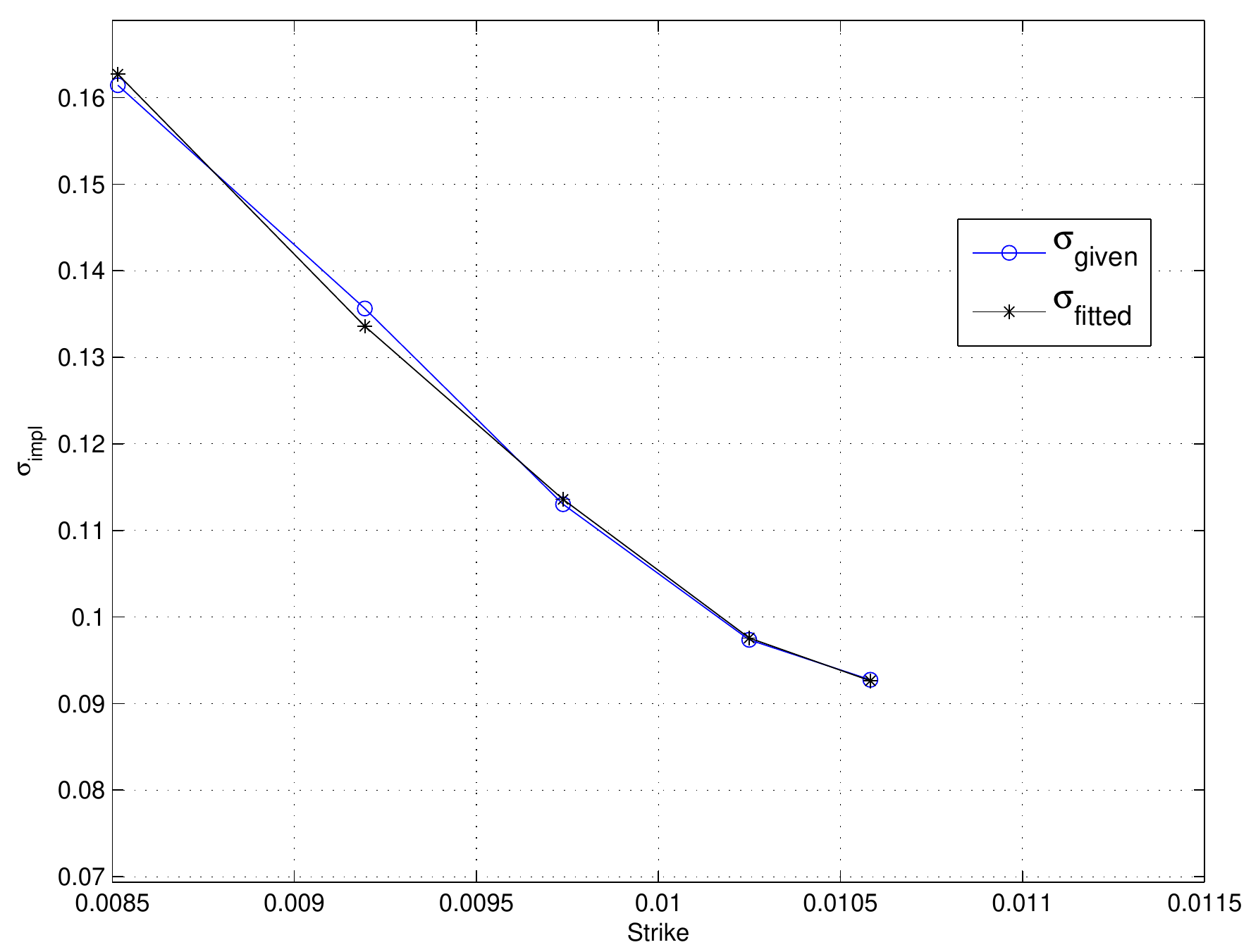}
\\[-12pt]\caption{fit for maturity 6M}
\label{fig:Heston_impliedVola_error_T_6M}
\end{minipage}\\[20pt]

\begin{minipage}[b]{0.45\textwidth}
\centering
\includegraphics[width=\textwidth]{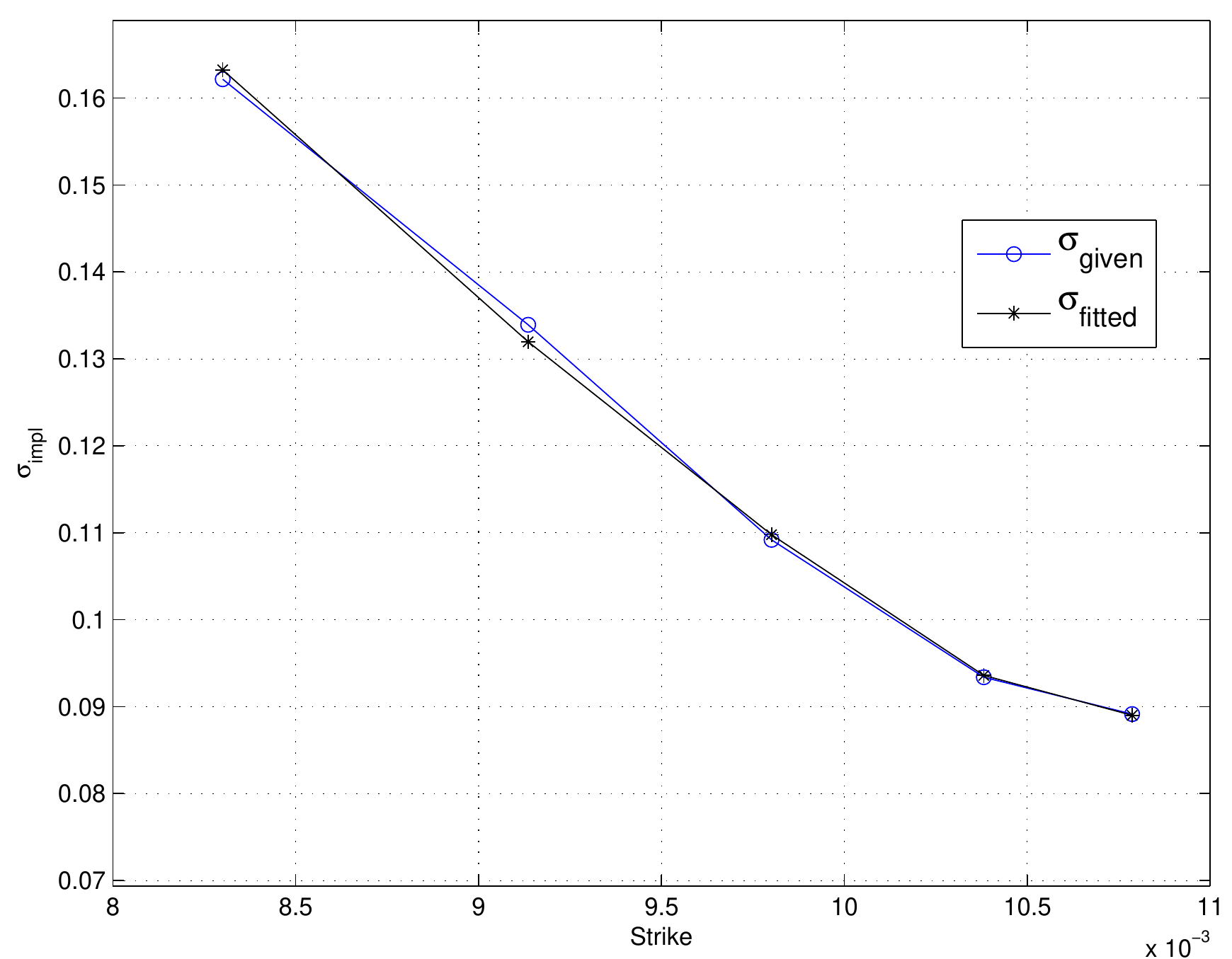}
\\[-12pt]\caption{fit for maturity 9M}
\label{fig:Heston_impliedVola_error_T_9M}
\end{minipage}%
\hspace{0.04\textwidth}%
\begin{minipage}[b]{0.45\textwidth}
\centering
\includegraphics[width=\textwidth]{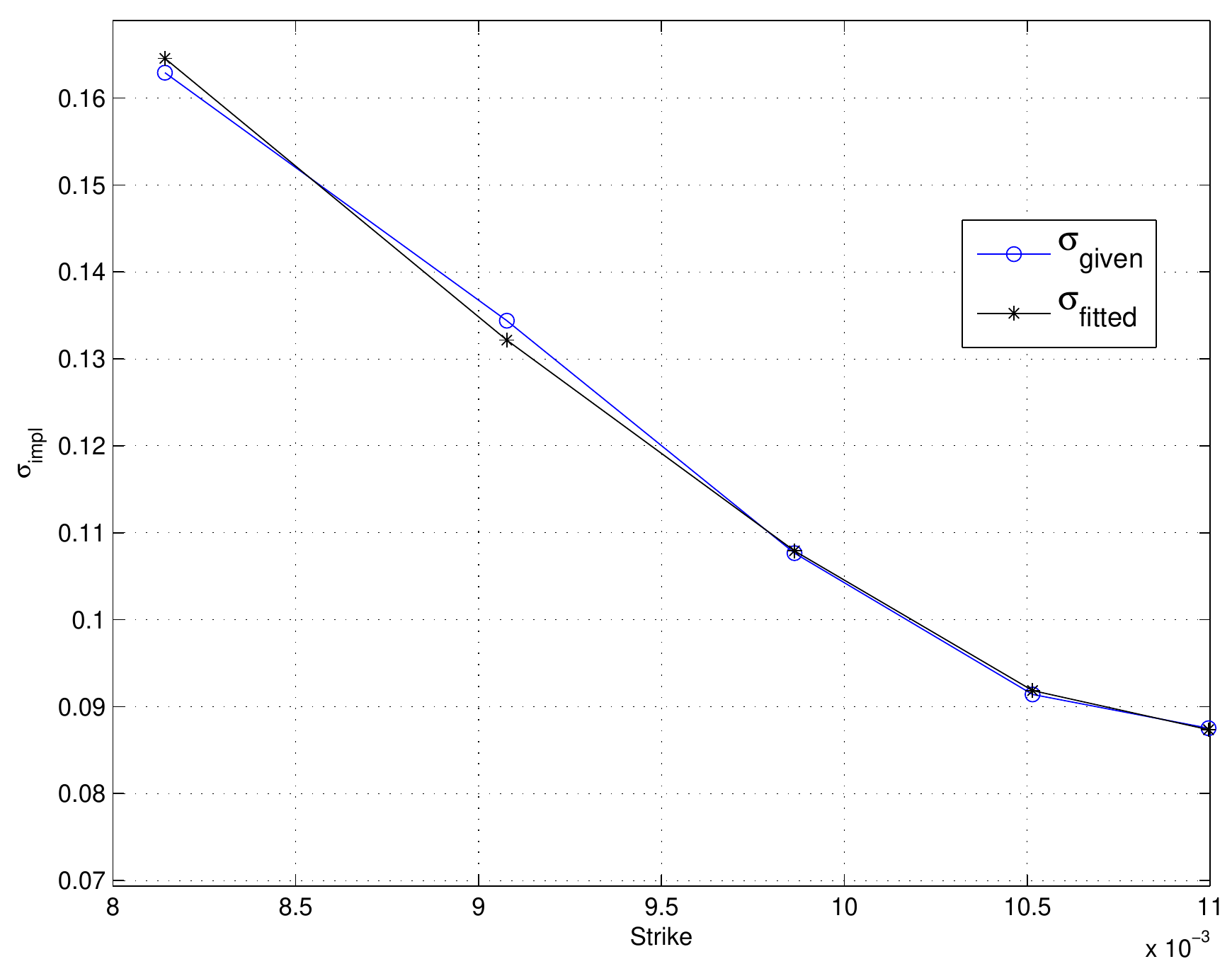}
\\[-12pt]\caption{fit for maturity 1Y}
\label{fig:Heston_impliedVola_error_T_1Y}
\end{minipage}\\[20pt]

\begin{minipage}[b]{0.45\textwidth}
\centering
\includegraphics[width=\textwidth]{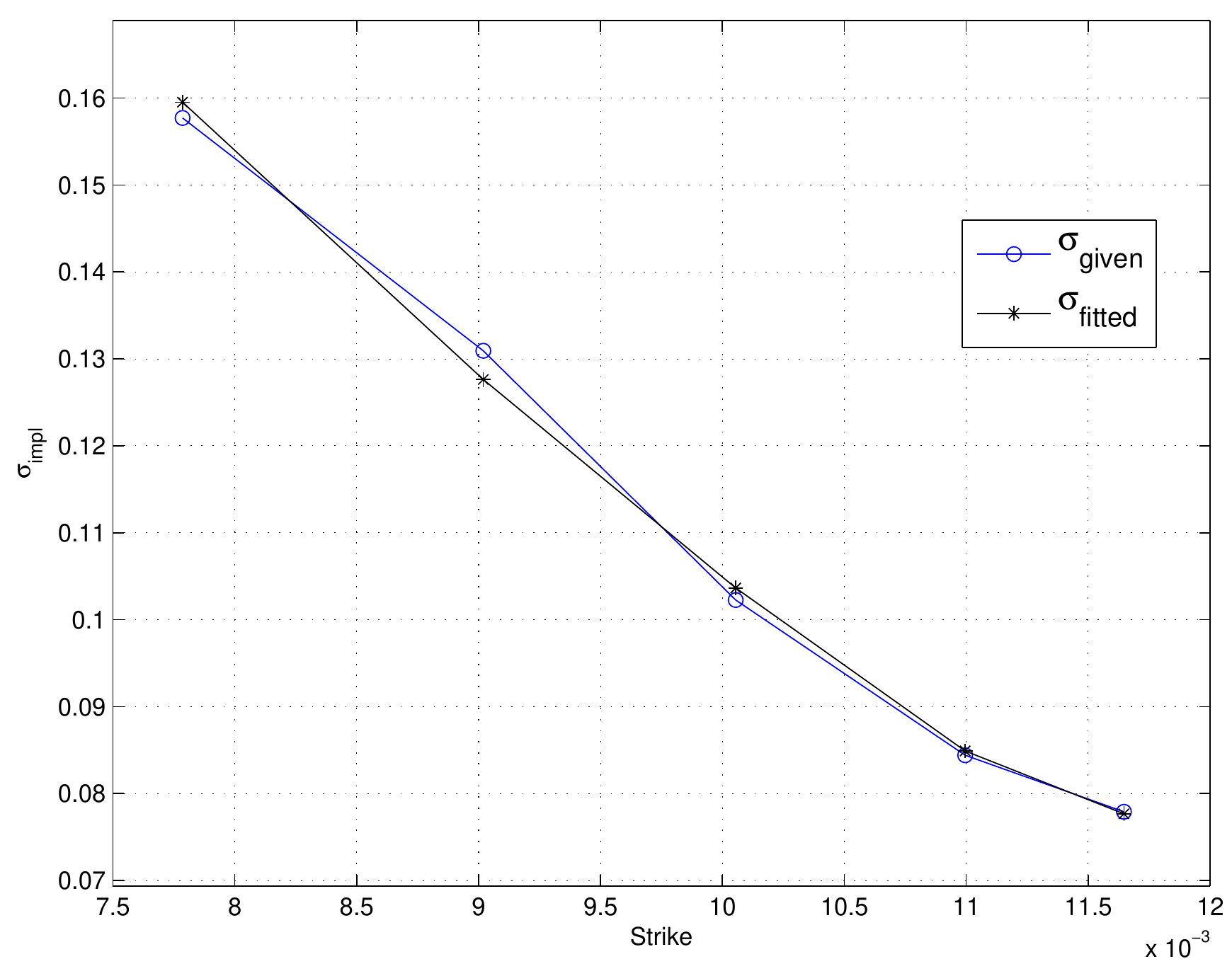}
\\[-12pt]\caption{fit for maturity 2Y}
\label{fig:Heston_impliedVola_error_T_2Y}
\end{minipage}%
\hspace{0.04\textwidth}%
\begin{minipage}[b]{0.45\textwidth}
\centering
\includegraphics[width=\textwidth]{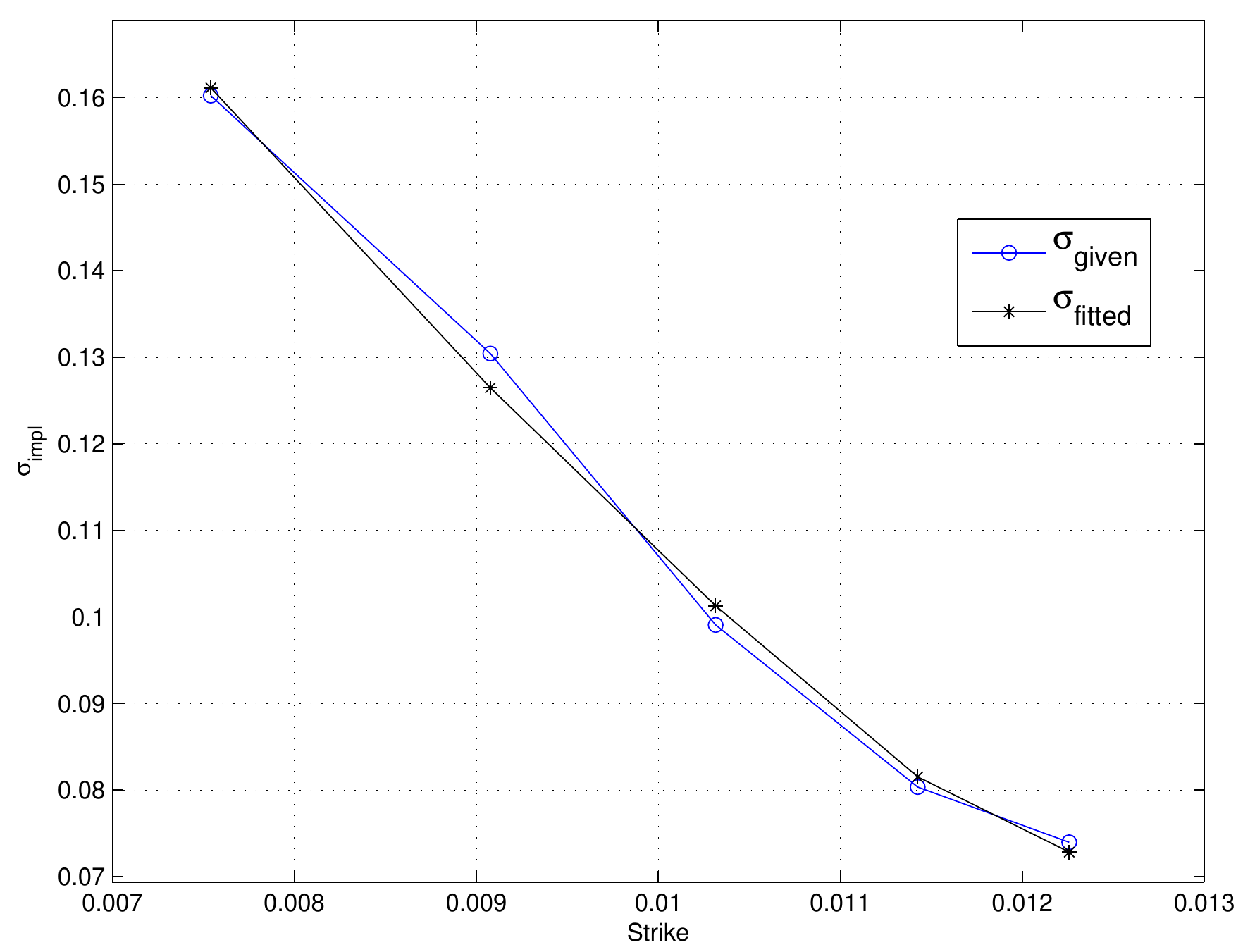}
\\[-12pt]\caption{fit for maturity 3Y}
\label{fig:Heston_impliedVola_error_T_3Y}
\end{minipage}

\end{figure}

\figref{fig:Heston_impliedVola_given} shows the given volatility surface quoted in terms of Deltas and \figref{fig:Heston_impliedVola_fitted} is the surface produced by our fitted model. \figref{fig:Heston_impliedVola_error} gives the difference between the quoted surface and the fitted surface. \figref{fig:Heston_time_dependent_params} illustrates how the parameters evolve over time. Due due the fact that the calibration is an ill posed problem, i.e.\ the solution might not be unique and the parameters don't depend continuously on the data,  we see considerable jumps of the time dependent parameters. \figref{fig:Heston_impliedVola_error_T_3M}--\ref{fig:Heston_impliedVola_error_T_3Y} show the given smile and the fitted smile for the various maturities.

\section{Conclusion and Outlook}
\label{sec:Conclusion}

We get a good fit of the given volatility surface for stochastic volatility models using piecewise constant parameters using a hybrid optimization algorithm and can use this fit to price exotic derivatives. We also note that the piecewise constant parameters may jump considerably. This is due to the fact the the calibration problem per se is ill posed giving evidence to the necessity of regularization techniques.
% Techniques like relative-entropy minimization may be used to improve the valuation of the options further, see \cite{AFHS97}.  
If the posterior distribution from parameter estimation from historical data is available, we may apply regularization techniques. Further, the bounds for the parameters for the optimization algorithm could be derived from (quantiles of) the posterior distribution and also the initial population could be sampled from the posterior distribution.

% \newpage
% BIBLIOGRAPHY -----------------------------------------------------------
\bibliographystyle{plainnat}
\begin{small}      
\bibliography{bibliography}
\end{small}
% ------------------------------------------------------------------------

\typeout{#####################################}
\end{document}